\documentclass[prd,twocolumn]{revtex4}
\usepackage{amssymb}
\usepackage{graphicx}
\usepackage{dcolumn}
\usepackage{bm}
\usepackage{amsfonts}
\usepackage{latexsym}
\usepackage{amsmath}
\usepackage{psfrag}
\usepackage[cp1251]{inputenc}
\usepackage[dvips]{epsfig}

\begin{document}
\title{A comment on ``Radiation reaction reexamined: bound momentum and
Schott term'' by D.V.~Gal'tsov and P.~Spirin}

\author{ P.O. Kazinski, S.L. Lyakhovich and A.A. Sharapov}
\affiliation{Department of Physics, Tomsk State University, 634050
Tomsk, Russia \\
{\rm{e-mails: kpo@phys.tsu.ru, sll@phys.tsu.ru,
sharapov@phys.tsu.ru}}}

\maketitle

In recent e-print \cite{GaS}, Gal'tsov and Spirin discuss the
radiation reaction  in $d=4$ classical electrodynamics of point
charges. In particular, reexamining the well known regularization
procedure for the retarded Green's function, the authors cast
doubt on the validity of our previous results from the paper
\cite{KLS} where the Lorentz-Dirac equation was derived in $d>4$,
as well as from the paper \cite{KS} devoted to the radiation
reaction in $d=4$ dynamics of the massless point particle. They
claim

\vspace{2mm} \textit{`` ... the regularization proposed in
\cite{KLS} fails to reproduce the correct result already in the
case of four dimensions, so the validity of the equations derived
in \cite{KLS}, \cite{KS} (and of the regularization itself) is
questionable.'' }

\vspace{2mm}

\noindent The authors justify this statement in a peculiar way.
They \textit{incorrectly reproduce formula} (23) \textit{from our
paper} \cite{KLS} (or the same formula (19) in \cite{KS}) for the
derivative of the delta function on the semi-axis. The correct
expression from \cite{KLS}, \cite{KS} reads
\begin{equation}\label{dp}
    \delta'(s)=-\lim_{a\rightarrow +0}\frac{\partial}{\partial
    a}\frac {e^{-s/a}}{a}\,,\quad s\geq 0\,.
\end{equation}
In \cite{GaS} it is taken without the overall minus sign. Using
the incorrect formula, Gal'tsov and Spirin then industriously
proceed to derive (providing a lot of details, see relations
(38-42) in \cite{GaS} and related explanations) the radiation
reaction force in $d=4$ with the wrong sign, directly inherited
from the minus they miss in (\ref{dp}).  This is what appears to
them a ``disproof'' of the regularization we use and the equations
we get in $d> 4$.

It was a simple exercise to check the sign in the representation
for $\delta'(s)$ (\ref{dp})  (even though it is a common
knowledge, our paper reminds how such expressions are derived),
but Gal'tsov and Spirin did not choose to do that. It is probable
they automatically copied this formula from our earlier e-print
hep-th/0201046 where the overall sign factor had accidentally
dropped out from this elementary relation. Although this minor
inaccuracy had been corrected in the journal version published two
years ago \cite{KLS} (and which is referred to by Gal'tsov and
Spirin), we considered this sign factor so obvious that we did not
regard this misprint  as a sufficient reason for replacing the
paper in the arXiv. In the next paper \cite{KS}, published a year
ago and also cited (again in relation with the regularization of
the delta function derivative) by Gal'tsov and Spirin, the sign
was correct both in the journal and in the e-print version.

It would be appropriate to mention here that the expression for
the Lorentz-Dirac force in any even dimension first obtained in
our work \cite{KLS}, checks well with the particular result for
d=6 derived in \cite{BK} by a completely different method. And, of
course, our result reproduces the text-book answer for d=4.

Although the Gal'tsov-Spirin e-print contains no new results (it
just compares well known achievements of the former times with
each other), the authors promise to shed new light on the
radiation reaction problem in higher dimensions elsewhere by
applying regularizations they prefer to any other scheme: the
point splitting one and/or that based on the integration of the
energy-momentum flux through a hypersurface encircling the
particle's world-line. It might be further appropriate to note
that the result of the renormalization of singular integrals
involving generalized functions of one variable, like $\delta(s^2
)$ and its derivatives, can not depend on a particular choice of
the regularization \cite{GSh}, unlike the many variable case usual
in the quantum field theory. This leaves no room for ambiguity in
the values like the radiation reaction force in classical
electrodynamics (see also the comments in Section 3 of our paper
\cite{KLS}). The regularization we used in \cite{KLS} has the
technical convenience because it is explicitly reparametrization
invariant and it allows to get the radiation reaction force in any
even dimension by a mere expansion of a simple generating function
in the regularization parameter. Another regularization can have
other technical advantages, but it will inevitably result in the
same equations as we get in \cite{KLS}.

\end{document}